# On Lines, Joints, and Incidences in Three Dimensions[*]

György Elekes[†]    Haim Kaplan[‡]    Micha Sharir[§]

October 31, 2018

This paper is dedicated to the memory of György Elekes, who passed away in September, 2008.

**Abstract**

We extend (and somewhat simplify) the algebraic proof technique of Guth and Katz [7], to obtain several sharp bounds on the number of incidences between lines and points in three dimensions. Specifically, we show: (i) The maximum possible number of incidences between $n$ lines in $\mathbb{R}^3$ and $m$ of their joints (points incident to at least three non-coplanar lines) is $\Theta(m^{1/3}n)$ for $m \geq n$, and $\Theta(m^{2/3}n^{2/3} + m + n)$ for $m \leq n$. (ii) In particular, the number of such incidences cannot exceed $O(n^{3/2})$. (iii) The bound in (i) also holds for incidences between $n$ lines and $m$ arbitrary points (not necessarily joints), provided that no plane contains more than $O(n)$ points and each point is incident to at least three lines. As a preliminary step, we give a simpler proof of (an extension of) the bound $O(n^{3/2})$, established by Guth and Katz, on the number of joints in a set of $n$ lines in $\mathbb{R}^3$. We also present some further extensions of these bounds, and give a proof of Bourgain's conjecture on incidences between points and lines in 3-space, which constitutes a simpler alternative to the proof of [7].

## 1   Background

In this paper we consider several extended variants of the problem of bounding the number of incidences between $n$ lines in $\mathbb{R}^3$ and their *joints*, where a joint is a point which is incident to (at least) three non-coplanar lines.

This problem extends a more basic one, of just bounding the number of joints. The latter problem has been around for almost 20 years [2, 6, 10], and, until very recently, the best known upper bound, established by Sharir and Feldman [6], was $O(n^{1.6232})$. The proof techniques were rather complicated, involving a battery of tools from combinatorial geometry, including forbidden subgraphs in extremal graph theory, space decomposition techniques, and some basic results in the geometry of lines in space (e.g., Plücker coordinates).

[*]Work by Haim Kaplan was partially supported by the United states - Israel Binational Science Foundation, project number 2006204, and by the Israel Science Foundation grant 975-06. Work by Micha Sharir has been supported by NSF Grants CCF-05-14079 and CCF-08-30272, by a grant from the U.S.-Israeli Binational Science Foundation, by grant 155/05 from the Israel Science Fund, and by the Hermann Minkowski–MINERVA Center for Geometry at Tel Aviv University.

[†]Department of Computer Science, Eötvös University, H-1117 Budapest, Hungary

[‡]School of Computer Science, Tel Aviv University, Tel Aviv 69978 Israel; *haimk@post.tau.ac.il*.

[§]School of Computer Science, Tel Aviv University, Tel Aviv 69978 Israel and Courant Institute of Mathematical Sciences, New York University, New York, NY 10012, USA; *michas@post.tau.ac.il*.

On the other hand, a simple construction, using the axis-parallel lines in a $k \times k \times k$ grid, for $k = \Theta(n^{1/2})$, has $3k^2 = \Theta(n)$ lines and $k^3 = \Theta(n^{3/2})$ joints. Notice that the number of incidences between the lines and joints in this construction is also $\Theta(n^{3/2})$, as every joint is incident to exactly three lines.

It has long been conjectured that the correct upper bound on the number of joints is $O(n^{3/2})$, matching the lower bound just noted. In a rather dramatic recent development, Guth and Katz [7] have settled the conjecture in the affirmative, showing that the number of joints is indeed $O(n^{3/2})$. Their proof technique is completely different, and uses fairly simple tools from algebraic geometry. As a preliminary step in our analysis, we will present a somewhat simplified version of their proof, in a more general context (see Section 3 for details).

The problem of bounding the number of line-joint incidences has also been studied; the most significant result to date is due to Sharir and Welzl [11], who established an upper bound of $O(n^{5/3})$ for this number. In an unpublished work, Elekes has shown that the number of incidences between $n$ *equally inclined* lines (lines forming a fixed angle with the $z$-axis) and their joints is $O(n^{3/2}\sqrt{\log n})$.

In this paper we extend the algebraic machinery of Guth and Katz, to show that the number of incidences between $n$ arbitrary lines in 3-space and their joints is $O(n^{3/2})$; as just noted, this bound is tight in the worst case. As a matter of fact, we obtain the following stronger results.

(i) The maximum possible number of incidences between $n$ lines in $\mathbb{R}^3$ and $m$ of their joints is $\Theta(m^{1/3}n)$ for $m \geq n$, and $\Theta(m^{2/3}n^{2/3} + m + n)$ for $m \leq n$. Since $m$ is at most $O(n^{3/2})$, this implies the $m$-independent bound mentioned above, namely $O(n^{3/2})$, on line-joint incidences.

(ii) The bound in (i) also holds for incidences between $n$ lines and $m$ arbitrary points (not necessarily joints), provided that no plane contains more than $O(n)$ points and each point is incident to at least three lines. It is easily checked that both conditions hold in the case of joints. As a preliminary step in the proof, we will show that the maximum number of points in this more general context is also $O(n^{3/2})$.

We also present some further extensions and consequences of these bounds.

Finally, we give an alternative (and, in our opinion, simpler) proof, to the one given by Guth and Katz [7], of a conjecture of Bourgain on incidences between points and lines in 3-space. This conjecture (now a theorem), inspired by Bourgain's work on Kakeya's problem (see, e.g., the survey paper of Tao [14] for details concerning Kakeya's problem and its connection to geometric incidence problems), is as follows: Given a set $L$ of $n$ lines in 3-space, and a set $P$ of points, such that[1] (i) no plane contains more than $n^{1/2}$ lines of $L$, and (ii) each line of $L$ contains at least $n^{1/2}$ points of $P$, then $|P| = \Omega(n^{3/2})$. A recent paper by Solymosi and Tóth [12] gave the weaker bound $|P| = \Omega(n^{11/8})$, before the conjecture was settled in the affirmative in [7].

We regard the present paper as a further opening of the door of combinatorial geometry to the new algebraic techniques, and it is our hope (and belief) that there will be many forthcoming applications of the new machinery to a variety of additional hard problems in the area. As a matter of fact, in work in progress, we have already managed to extend the new ideas to another incidence problem, involving points and a special class of parabolas in $\mathbb{R}^3$, which is strongly related to the problem of distinct distances in the plane.

---

[1]The parameter $n^{1/2}$ appearing (twice) in the assumptions can be replaced by any constant multiples of $n^{1/2}$.

## 2 Tools from algebraic geometry

We begin by reviewing, extending, and somewhat simplifying the basic tools from algebraic geometry which have been used in [7].

First, note that a trivariate polynomial $p$ of degree $d$ which vanishes at $d+1$ collinear points must vanish identically on their supporting line.

**Critical points and lines.** A point $a$ is *critical* (or *singular*) for a trivariate polynomial $p$ if $p(a) = 0$ and $\nabla p(a) = 0$; any other point $a$ in the zero set of $p$ is called *regular*. A line $\ell$ is *critical* if all its points are critical.

**Proposition 1** *Let $p$ and $p'$ be two trivariate polynomials of respective degrees $k$ and $m$, such that $p$ and $p'$ have no common factors. Then there are at most $km$ lines on which both $p$ and $p'$ vanish identically.*

**Proof.** Assume that $p$ and $p'$ vanish identically on $km + 1$ lines and that, without loss of generality, none of these lines is parallel to the plane $z = 0$. Then, for every $c$, all these lines intersect the plane $z = c$ transversally. This implies, by Bézout's theorem [3, 4], that $p$ and $p'$, as bivariate polynomials restricted to $z = c$, have a common factor. Since this holds for every $c$, it follows that $p$ and $p'$ themselves, as trivariate polynomials, have a common factor, which is a contradiction. (To see the latter claim, assume, without loss of generality, that both $p$ and $p'$ have positive degrees in $x$—a random rotation of the coordinate frame about the $z$-axis will ensure this property. Then the *resultant* of $p$ and $p'$ (see, e.g., [4]), as polynomials in $x$ (this resultant is a polynomial in $y$ and $z$) vanishes identically on the plane $z = c$ for every $c$. This means that the resultant of $p$ and $p'$ vanishes identically in $\mathbb{R}^3$ as a polynomial in $y$ and $z$, and so $p$ and $p'$ have a common factor.) □

Since the components of $\nabla p$ are three polynomials of degree $d-1$ we obtain the following immediate corollary of Proposition 1, by applying it to $p$ and to any of its partial derivatives.

**Corollary 2** *An irreducible trivariate polynomial $p$ of degree $d$ can have at most $d(d-1)$ critical lines.*

We next show that irreducibility of $p$ is not really needed.

**Proposition 3** *Any trivariate square-free polynomial $p$ of degree $d$ can have at most $d(d-1)$ critical lines.*

**Proof.** We prove the claim by induction on the degree $d$ of $p$. The claim holds trivially for $d = 1$, so assume that $d > 1$.

If $p$ is irreducible, the claim is established in Corollary 2. Suppose then that $p$ is reducible, and write $p = fg$, so that $f$ and $g$ are nonconstant square-free polynomials with no common factor (since $p$ is square-free, this can always be done). Denote the degrees of $f$ and $g$ by $d_f$ and $d_g$, respectively; we have $d_f, d_g \geq 1$ and $d = d_f + d_g$.

Let $\ell$ be a critical line for $p$. Then either $f \equiv 0$ on $\ell$ or $g \equiv 0$ on $\ell$ (or both). Moreover, since $\nabla p = f\nabla g + g\nabla f \equiv 0$ on $\ell$, it is easily checked that $\ell$ must satisfy (at least) one of the following properties:

(i) $f \equiv g \equiv 0$ on $\ell$.

(ii) $\ell$ is a critical line of $f$.

(iii) $\ell$ is a critical line of $g$.

Indeed, if (i) does not hold, we have, without loss of generality, $f \equiv 0$ on $\ell$, but $g$ vanishes only at finitely many points of $\ell$. On any other point $a$ of $\ell$ we then must have $\nabla f(a) = 0$, which implies that $\nabla f$ is identically zero on $\ell$, so $\ell$ is critical for $f$. This implies (ii); (iii) holds in the symmetric case where $g \equiv 0$ on $\ell$ but $f$ does not vanish identically on $\ell$.

By the induction hypothesis, the number of critical lines for $f$ is at most $d_f(d_f - 1)$, and the number of critical lines for $g$ is at most $d_g(d_g - 1)$. By Proposition 1, at most $d_f d_g$ lines satisfy (i). Altogether, the number of critical lines for $p$ is at most

$$d_f(d_f - 1) + d_g(d_g - 1) + d_f d_g < d(d-1).$$

□

**Proposition 4** *Let $a$ be a regular point of $p$, so that $p$ vanishes at three lines passing through $a$. Then these lines must be coplanar.*

**Proof.** Any such line must be contained in the tangent plane to $p = 0$ at $a$. □

Hence, a point $a$ incident to three non-coplanar lines on which $p$ vanishes must be a critical point of $p$.

**Proposition 5** *Given a set $S$ of $m$ points in 3-space, there exists a trivariate polynomial $p(x, y, z)$ which vanishes at all the points of $S$, whose degree is at most the smallest integer $d$ satisfying $\binom{d+3}{3} > m$.*

**Proof.** A trivariate polynomial of degree $d$ has $\binom{d+3}{3}$ monomials, and requiring it to vanish at $m < \binom{d+3}{3}$ points yields $m$ linear homogeneous equations in the coefficients of these monomials. Such an underdetermined system always has a nontrivial solution. □

**Flat points and lines.** Call a regular point $a$ of a trivariate polynomial $p$ *linearly flat* if it is incident to three distinct (necessarily coplanar) lines on which $p$ vanishes identically.

Let $a$ be a linearly flat point of $p$, and let $\ell_1, \ell_2, \ell_3$ be three incident lines on which $p$ vanishes. The second-order Taylor expansion of $p$ at $a$ has the form

$$\begin{aligned} q(u) &= p(a) + \nabla p(a) \cdot (u - a) + \frac{1}{2}(u-a)^T H_p(a)(u-a) \\ &= \nabla p(a) \cdot (u-a) + \frac{1}{2}(u-a)^T H_p(a)(u-a), \end{aligned}$$

where
$$H_p = \begin{pmatrix} p_{xx} & p_{xy} & p_{xz} \\ p_{xy} & p_{yy} & p_{yz} \\ p_{xz} & p_{yz} & p_{zz} \end{pmatrix}$$

is the *Hessian* of $p$. $q$ is a quadratic polynomial (in $u$) which approximates $p$ up to third-order terms for $u$ sufficiently close to $a$. Hence, substituting $u = a + \varepsilon v_i$, where $v_i$ is the direction of $\ell_i$, for $i = 1, 2, 3$, and $\varepsilon$ is sufficiently small, we get

$$0 = \varepsilon \nabla p(a) \cdot v_i + \frac{1}{2}\varepsilon^2 v_i^T H_p(a) v_i + O(\varepsilon^3),$$

so we must have, for each $i$,

$$\nabla p(a) \cdot v_i = v_i^T H_p(a) v_i = 0.$$

This in turn implies that $q$ vanishes *identically* on each of the lines $\ell_i$. Then any line $\ell$ in the tangent plane $\pi_a$ of $p = 0$ at $a$, not incident to $a$ and not parallel to any $\ell_i$, intersects each of the lines $\ell_i$ at a distinct point, so $q$ vanishes on three distinct points of $\ell$, and thus, being a quadratic polynomial, it vanishes identically on $\ell$, and thus on $\pi_a$.

For example, assume that $a$ is the origin and that $\pi_a$ is the $xy$-plane. Then the condition that $q$ vanishes on $\pi_a$ is equivalent to

$$p(a) = p_{xx}(a) = p_{xy}(a) = p_{yy}(a) = 0 \tag{1}$$

(because $q$ becomes the identically zero polynomial when substituting $z = 0$). To make this condition independent of the orientation of $\pi_a$, we note that $q$ has to vanish at any $u$ such that $u - a$ is orthogonal to $\nabla p(a)$. We claim that $q \equiv 0$ on $\pi_a$ if and only if $q$ vanishes at the three points

$$u_j = a + \nabla p(a) \times e_j,$$

where $e_1, e_2, e_3$ are the standard coordinate unit vectors. (We assume here general position of the coordinate frame, so that $\nabla p(a)$ is not parallel to any coordinate plane at any of the finitely many points $a$ that we will consider later. Under this assumption, the three vectors $u_j$ are distinct, at each such point $a$.) Indeed, necessity of this condition is clear, and sufficiency is proved as in the earlier part of the argument. That is, if $q$ vanishes at $u_j$, it vanishes identically on the line through $a$ and $u_j$. This follows from the fact that the first-order part of $q$ vanishes at $u_j$, so the second-order part also vanishes. This implies that $q$ vanishes at each point of the form $a + t \nabla p(a) \times e_j$, for any $t \in \mathbb{R}$, that is, $q$ vanishes on the line through $a$ and $u_j$. Since this holds for $j = 1, 2, 3$, the preceding argument implies that $q \equiv 0$ on $\pi_a$. Since the first-order component of $q$ vanishes "automatically" on such vectors $u$, the condition is equivalent to

$$(\nabla p(a) \times e_j)^T H_p(a)(\nabla p(a) \times e_j) = 0, \quad \text{for } j = 1, 2, 3.$$

This suggests that we define the three polynomials[2]

$$\Pi_j(p)(u) = (\nabla p(u) \times e_j)^T H_p(u)(\nabla p(u) \times e_j), \quad \text{for } j = 1, 2, 3, \tag{2}$$

which, as we have just shown, satisfy the following properties.

---
[2]Guth and Katz [7] consider instead the nine polynomials $(\nabla p(u) \times e_i)^T H_p(u)(\nabla p(u) \times e_j)$, for $i, j = 1, 2, 3$, but the three that we use seem to suffice.

**Proposition 6** *Let $p$ be a trivariate polynomial, and let $a$ be a regular point of $p$. (i) If $a$ is linearly flat then $\Pi_j(p)(a) = 0$, for $j = 1, 2, 3$. (ii) Conversely, if $\Pi_j(p)(a) = 0$, for $j = 1, 2, 3$, then $a$ is "quadratically flat," in the sense that the quadratic second-order Taylor approximation $q$ vanishes identically on the tangent plane $\pi_a$ of $p$ at $a$.*

Note that if the degree of $p$ is $d$ then each $\Pi_j(p)$ is a polynomial of degree at most $(d-1) + (d-2) + (d-1) = 3d - 4$.

In differential geometry parlance (see, e.g., [9]) the property in part (ii) of the theorem can be restated as follows. The *second fundamental form* of the zero set $Z$ of $p$ at a regular point $a$ is defined as $A du^2 + 2B du dv + C dv^2$, where $\mathbf{x} = \mathbf{x}(u, v)$ is a parametrization of $Z$ (locally near $a$), and

$$A = \mathbf{x}_{uu} \cdot \mathbf{n}, \qquad B = \mathbf{x}_{uv} \cdot \mathbf{n}, \qquad C = \mathbf{x}_{vv} \cdot \mathbf{n},$$

where $\mathbf{n} = \nabla p(a)/\|\nabla p(a)\|$ is the unit normal to $Z$ at $a$.

Then, as is easily verified, property (ii) holds at a regular point $a \in Z$ if and only if the second fundamental form vanishes at $a$ (i.e., $A = B = C = 0$ at $a$).

In what follows, we call a point $a$ *flat* for $p$ if $\Pi_j(p)(a) = 0$, for $j = 1, 2, 3$, or, equivalently, if the second fundamental form of $p$ vanishes at $a$.

Call a line $\ell$ *flat* for $p$ if all the points of $\ell$ are flat points of $p$ (with the possible exception of finitely many critical points). Clearly, if $\ell$ contains at least $3d - 3$ flat points then $\ell$ is a flat line.

Next, we show that, in general, trivariate polynomials do not have too many flat lines. As before, we first establish this property for irreducible polynomials, and then extend the analysis to more general polynomials.

**Proposition 7** *Let $p$ be an irreducible trivariate polynomial of degree $d > 1$. Then $p$ can have at most $3d^2 - 4d$ flat lines.*

**Proof.** Suppose to the contrary that there are more than $3d^2 - 4d$ flat lines. By Proposition 1, $p$ and $\Pi_1(p)$ must have a common factor. Since $p$ is irreducible, $p$ must be a factor of $\Pi_1(p)$. Similarly, $p$ must be a factor of $\Pi_2(p)$ and of $\Pi_3(p)$. This implies that all the (regular) points at which $p$ vanishes are flat.

By Proposition 6(ii) and the remarks following that proposition, it follows that the second fundamental form of the zero set $Z$ vanishes at all the regular points of $p$. This in turn implies that $Z$, locally near any regular point, is a portion of a plane. This is a well known result—see, e.g., Exercise 6.2 in [9]. For the sake of completeness, we include a short proof in an appendix. This property, and the irreducibility of $p$, imply that $Z$ is a plane, or that $p$ is a linear polynomial, contradicting our assumption that its degree is greater than 1. □

**Proposition 8** *Let $p$ be any trivariate square-free polynomial of degree $d$ with no linear factors. Then $p$ can have at most $3d^2 - 4d$ flat lines.*

**Proof.** We proceed by induction on the degree of $p$, where the basis of the induction, for $d = 2$, holds because in this case $p$, which has no linear factors, must be irrducible, and the claim follows from Proposition 7.

Assume then that $p$ has degree $d \geq 3$. If $p$ is irreducible, the claim holds by Proposition 7. Otherwise, write $p = fg$ where $f$ and $g$ are nonconstant square-free polynomials with no common factors and no linear factors. Let $d_f$ and $d_g$ denote their respective degrees, so $d = d_f + d_g$.

We claim that any flat point of $p$, at which only one of $f, g$ vanishes, must be a flat point of the respective vanishing factor ($f$ or $g$). Indeed, consider a point $a$ where $f$ vanishes but $g$ does not. We have

$$\nabla p(a) = f(a)\nabla g(a) + g(a)\nabla f(a) = g(a)\nabla f(a),$$

and, for any vector $u$, one can easily verify that

$$u^T H_p(a) u = g(a)(u^T H_f(a) u) + (\nabla f(a) \cdot u)(\nabla g(a) \cdot u).$$

Hence, substituting

$$u = \nabla p(a) \times e_j = g(a)(\nabla f(a) \times e_j),$$

we get $\Pi_j(p)(a) = g^3(a)\Pi_j(f)(a)$, from which the claim follows.

By Proposition 1, there are at most $d_f d_g$ lines on which both $f$ and $g$ vanish identically and simultaneously. Hence, any other flat line for $p$ must be a flat line for either $f$ or $g$. By induction, $f$ has at most $3d_f^2 - 4d_f$ flat lines and $g$ has at most $3d_g^2 - 4d_g$ flat lines. Summing up the number of critical lines of all types, we get the bound

$$3d_f^2 - 4d_f + 3d_g^2 - 4d_g + d_f d_g < 3d^2 - 4d,$$

and the lemma follows. □

## 3 The Guth–Katz bound: Review and extension

In preparation for the proof of our main results, we review and extend the proof of Guth and Katz [7], to obtain the following result.

**Theorem 9** *Let $L$ be a set of (at most) $n$ lines in $\mathbb{R}^3$ and let $P$ be a set of $m$ arbitrary points in $\mathbb{R}^3$, such that (i) no plane contains more than $bn$ points of $P$, for some absolute constant $b \geq 1$, and (ii) each point of $P$ is incident to at least three lines of $L$. Then $m = O(n^{3/2})$ (where the constant of proportionality depends linearly on $b$).*

**Remark.** Let $J = J_L$ denote the set of the joints of $L$, namely, points incident to (at least) three non-coplanar lines of $L$. Guth and Katz [7] show that $|J| = O(n^{3/2})$. This is a special case of Theorem 9, because the set $J_L$ of joints of $L$ satisfies the conditions of this theorem. Indeed, condition (ii) is obvious. For condition (i), consider some plane $\pi$. Every point $a \in J_L \cap \pi$ must be incident to at least one line of $L$ which is not contained in $\pi$, and each such line intersects $\pi$ at a unique point, so $\pi$ cannot contain more than $n$ points of $J_L$.

We give a proof of the more general Theorem 9, which, in our opinion, is simpler than the proof of Guth and Katz, although it draws heavily on their ideas.[3] It proceeds as follows.

**Proof.** We prove, using induction on the number of lines, that $|P| \leq An^{3/2}$, for some sufficiently large absolute constant $A$, whose choice will be dictated by several constraints that will arise during the proof. Clearly, $|P| < n^2$, so, by choosing $A > n_0^{1/2}$, for some sufficiently large constant $n_0$, the theorem trivially holds for any $n \leq n_0$, thus establishing the base of the induction.

For the induction step, assume that the theorem holds for all sets $L'$ and $P'$, as specified, where $|L| < n$. Consider a set $L$ of $n$ lines and a corresponding set $P$ of points satisfying the assumptions of the theorem, and suppose to the contrary that $|P| > An^{3/2}$.

**Pruning.** We first apply the following iterative pruning process to $L$. As long as there exists a line $\ell \in L$ incident to fewer than $cn^{1/2}$ points, for some constant $c \ll A$ that we will fix later, we remove $\ell$ from $L$, remove its incident points from $P$, and repeat this step with respect to the reduced sets of lines and points (keeping the threshold $cn^{1/2}$ fixed). In this process we delete at most $cn^{3/2}$ points. We are thus left with a subset of the original lines, each incident to at least $cn^{1/2}$ surviving points, and each surviving point is incident to at least three surviving lines. For simplicity, continue to denote these sets as $L$ and $P$. Let $n_1$ denote the number of lines left in $L$ after the pruning.

**Sampling.** Choose a random sample $L^s$ of lines of $L$, by picking each line of $L$ independently with probability $t$, where $t < 1$ is a small positive constant that we will fix later.

The expected number of lines which we choose is $tn_1 \leq tn$. Consider a line $\ell \in L \setminus L_s$. Since each point $a \in P \cap \ell$ is incident to a line of $L^s$ with probability at least $t$, the expected number of points in $P \cap \ell$ which lie on lines of $L^s$ is at least $ctn^{1/2}$. Hence, using Chernoff's bound (see, e.g., [1]) and the probability union bound, we obtain that, with positive probability, (a) $\frac{1}{2}tn_1 \leq |L^s| \leq 2tn_1$. (b) Each line $\ell \in L$ contains at least $\frac{c}{2}tn^{1/2}$ points that lie on lines of $L^s$.

Indeed, each surviving line is incident to at least $cn^{1/2}$ surviving points, each incident to at least two distinct other surviving lines, so we must have $n_1 \geq 2cn^{1/2}$. This means that the failure probability of the event specified by (a) is polynomially small in $n$, and this is obviously also the case for each of the $O(n)$ events specified in (b). The union bound then implies that the probability that either (a) or (b) fails is bounded by a constant.

We assume that $L^s$ does indeed satisfy (a) and (b), and choose $n^{1/2}$ arbitrary points on each line in $L^s$, to obtain a set $S$ of at most $2tn^{3/2}$ points. We guarantee that $S$ is not empty by choosing $t$ and $c$ such that $\frac{1}{2}tn_1 > tc > 1$.

**The polynomial $p$.** Applying Proposition 5, we obtain a polynomial $p(x, y, z)$ which vanishes at all the points of $S$, whose degree is at most the smallest integer $d$ satisfying

---

[3]To be fair, the proof of Guth and Katz, catering only to the case of joints, does not have to use the machinery involving flat points and lines, which we have to use, since we handle a more general situation. (They develop this machinery only for the proof of Bourgain's conjecture.) Nevertheless (as we feel), with the availability of this machinery, our proof is still simpler.

$\binom{d+3}{3} \geq |S| + 1$, so the degree is at most

$$d \leq \lceil (6|S|)^{1/3} \rceil \leq 2(12t)^{1/3} n^{1/2},$$

for $n_0$ (and thus $n$) sufficiently large. Without loss of generality, we may assume that $p$ is square-free: by removing repeated factors, we get a square-free polynomial which vanishes on the same set as the original $p$, with the same upper bound on its degree.

The polynomial $p$ vanishes on $n^{1/2}$ points on each line in $L^s$. This number is larger than $d$, if we choose $t$ sufficiently small so as to satisfy $2(12t)^{1/3} < 1$. Hence $p$ vanishes identically on the lines in $L^s$. Any other line of $L$ meets at least $\frac{c}{2} t n^{1/2}$ lines of $L^s$, and we can make this number also larger than $d$, with an appropriate choice of $t$ and $c$ (we need to ensure that $\frac{c}{2} t > 2(12t)^{1/3}$). Hence, $p$ vanishes on each line of $L$. We will also later need the property that each line of $L$ contains at least $5d$ points of $P$; that is, we require that $cn^{1/2} > 5d$, which will hold if $c > 10(12t)^{1/3}$.

To recap, the preceding paragraphs impose several inequalities on $c$ (and thereby on $A$) and $t$, and a couple of additional inequalities will be imposed later on. All these inequalities are easy to satisfy by choosing $t < 1$ to be a sufficiently small positive constant, and $A$ a sufficiently large constant.

We note that $p$ can have at most $d$ linear factors; i.e., $p$ can vanish identically on at most $d$ planes $\pi_1, \ldots, \pi_k$, for $k \leq d$. We factor out all the linear factors from $p$, and let $\tilde{p}$ denote the resulting polynomial, which is a square-free polynomial without any linear factors, of degree at most $d$.

By assumption, each plane $\pi_i$ contains at most $bn$ points of $P$. So on all planes together we have at most $bnd \leq 2b(12t)^{1/3} n^{3/2}$ points. We remove these planes together with the points and the lines contained in them, and let $L_1 \subseteq L$ and $P_1 \subseteq P$ denote, respectively, the set of those lines of $L$ (points of $P$) which are not contained in any of the vanishing planes $\pi_i$.

Note that there are still at least three lines of $L_1$ incident to any remaining point in $P_1$, since none of the points of $P_1$ lies in any plane $\pi_i$, so all lines incident to such a point are in $L_1$.

Clearly, $\tilde{p}$ vanishes identically on every $\ell \in L_1$. Furthermore, every $\ell \in L_1$ contains at most $d$ points in the planes $\pi_i$. Hence, $\ell$ contains at least $4d$ points of $P_1$. Since each of these points is incident to at least three lines in $L_1$, each of these points is either critical or linearly flat for $\tilde{p}$.

Consider a line $\ell \in L_1$. If $\ell$ contains more than $d$ critical points then $\ell$ is a critical line for $\tilde{p}$. By Proposition 3, the number of such lines is at most $d(d-1)$. Any other line $\ell \in L_1$ contains more than $3d - 4$ linearly flat points and hence $\ell$ must be a flat line for $\tilde{p}$. By Proposition 8, the number of such lines is at most $d(3d - 4)$. Summing up we obtain

$$|L_1| = d(d-1) + d(3d-4) < 4d^2 < 16(12t)^{2/3} n < n,$$

where the last inequality can be enforced with an appropriate choice of $t$. (We have worked hard just to get the inequality $|L_1| < n$, which means that at least one line has been removed from $L$ either in the pruning stage or because it lies in some vanishing plane of $p$; of course, by choosing a smaller value for $t$, we can make the size of $L_1$ much smaller.)

We next want to apply the induction hypothesis to $L_1$ and $P_1$, using $4d^2$ as the bound on the size of $L_1$. For this, we first need to argue that no plane contains more than $4bd^2$ points of $P_1$. Indeed, let $\pi$ be an arbitrary plane. Note first that $\pi$ contains at most $d$ lines of $L_1$, for otherwise $\tilde{p}$ would have vanished identically on $\pi$, contrary to the fact that $\tilde{p}$ has no linear factors. Each point of $P_1 \cap \pi$ is incident to at least three lines of $L_1$. Since $\pi$ contains no more than $d$ lines, at most $\binom{d}{2}$ of the points in $\pi$ are incident to at least two lines contained in $\pi$. Any other point in $P_1 \cap \pi$ is incident to at least two lines of $L_1$ not contained in $\pi$, and each such line intersects $\pi$ at a single point. Hence the number of points in $P_1 \cap \pi$ incident to at least two lines of $L_1$ that are not contained in $\pi$ is at most $4d^2/2 = 2d^2$. Together, $\pi$ contains at most $2d^2 + \binom{d}{2} < 4d^2 \leq 4bd^2$ points of $P_1$.

Hence, the lines in $L_1$ satisfy the assumption of the theorem for $4d^2 < n$. So, by induction, the number of points in $P_1$ is at most $A(4d^2)^{3/2} = 768tAn^{3/2}$. Adding up the bounds on the number of points on lines removed during the pruning process and in the planes $\pi_i$ (which correspond to the linear factors of $p$), we obtain

$$m \leq 768tAn^{3/2} + 2b(12t)^{1/3}n^{3/2} + cn^{3/2} \leq An^{3/2},$$

with an appropriate, final choice of $t$, $c$, and $A$. This contradicts the assumption on $P$ and $L$, and thus establishes the induction step and completes the proof of the theorem. □

## 4 Incidences between lines and points in $\mathbb{R}^3$

As in the previous section, let $L$ be a set of $n$ lines in $\mathbb{R}^3$, and let $J_L$ denote the set of their *joints*, namely, points incident to at least three non-coplanar lines of $L$.

In this section we further extend the proof technique of Guth and Katz [7] to obtain a bound on $I(J,L)$, the number of incidences between an arbitrary subset $J$ of $J_L$ and $L$. Specifically, we show:

**Theorem 10** *Let $L$ be a set of $n$ lines in $\mathbb{R}^3$ and let $J$ be a set of $m$ joints of $L$. Then*

$$I(J,L) = \min\left\{O(m^{1/3}n),\ O(m^{2/3}n^{2/3} + m + n)\right\}.$$

*The bound is tight in the worst case.*

As a matter of fact, we will establish a more general result, stated below, which will immediately imply the upper bound of Theorem 10. Specifically, we have:

**Theorem 11** *Let $L$ be a set of (at most) $n$ lines in $\mathbb{R}^3$ and let $P$ be a set of $m$ arbitrary points in $\mathbb{R}^3$, such that (i) no plane contains more than $bn$ points of $P$, for some absolute constant $b \geq 1$, and (ii) each point of $P$ is incident to at least three lines of $L$. Then*

$$I(P,L) = \min\left\{O(m^{1/3}n),\ O(m^{2/3}n^{2/3} + m + n)\right\}.$$

*Here too the bound is tight in the worst case.*

**Discussion.** (a) Note that the second expression in the bounds of the above theorems is simply the Szemerédi-Trotter bound on the number of incidences between $m$ points and $n$ lines in the plane [13]. This always serves as an upper bound for point-line incidences in 3-space (or in any higher dimension), since we can project the given points and lines onto some generic plane $\pi$, so that the projected points and lines are distinct, note that incidences are preserved by such a projection, and apply the planar bound of [13] to the projected points and lines. The novelty is in the first bound, which becomes smaller than the second bound when $m \geq n$.

(b) Note that Theorem 11 caters to the same situation considered in Theorem 9. It is interesting to note that, to obtain the improved bound $O(m^{1/3}n)$, we need both conditions assumed in the theorem: If the first condition is not imposed, we can construct $m$ points and $n$ lines, all lying in a common plane, so that each point is incident to at least three lines and so that the number of incidences between the points and lines is $\Theta(m^{2/3}n^{2/3} + m + n)$ (see, e.g., [5, 8]), which becomes larger than $O(m^{1/3}n)$ when $m > n$. If the second condition is not imposed, we can place all our points and lines on the hyperbolic paraboloid $z = xy$, so that half of the lines belong to one generating family and the other half belong to the second family, and take for $P$ the set of all their intersections. In this case $m = \Theta(n^2)$ and the number of incidences is also quadratic, and the bound $O(m^{1/3}n) = O(n^{5/3})$ does not hold. Note that in this construction condition (i) holds, because at most two lines in the construction (one from each family) can be coplanar, and the plane that they define contains at most $n$ points of $P$. The same constructions show, as is easily verified, that Theorem 9 may also fail if we drop any of its assumptions.

(c) Note that Theorem 11 implies Theorem 9, because, trivially, $I(P, L) \geq 3m$, from which we get $3m = O(m^{1/3}n)$, or $m = O(n^{3/2})$. However, the proof of Theorem 11 uses Theorem 9, so the latter theorem requires an independent proof, as provided in the preceding section.

**Proof of Theorem 10.** Most of the effort will be spent in the proof of Theorem 11. To clear the way for that proof, we first assume it to hold, and apply it to extablish Theorem 10.

**Upper bound.** The upper bound follows because $L$ and $J$ satisfy conditions (i) and (ii) of Theorem 11. This follows by the same argument as the one showing that Theorem 9 generalizes the result of Guth and Katz.

**Lower bound.** Consider first the case $m \leq n$. Construct $m$ points and $n$ lines in the $xy$-plane, say, with $\Theta(m^{2/3}n^{2/3} + m + n)$ incidences between them (see, e.g., [8]), such that each point is incident to at least two distinct lines, and pass an additional vertical line (in the $z$-direction) through each point. This yields $m + n \leq 2n$ lines and $m$ points, each of which is now a joint, with $\Theta(m^{2/3}n^{2/3} + m + n)$ incidences.

Consider next the case $m \geq n$. Put $t = m/n$. Construct in the $xy$-plane $n$ points and $n/t$ lines with $\Theta(n^{2/3}(n/t)^{2/3} + n)$ incidences between them. By Theorem 9, $t = O(n^{1/2})$, so $n/t = \Omega(n^{1/2})$. In this case, the first term in the bound dominates, and the number of incidences is thus $\Theta(n^{4/3}/t^{2/3})$. Now shift upwards the construction $t - 1$ times, into the parallel planes $z = 1, \ldots, z = t - 1$, to obtain a total of $tn = m$ points, $t(n/t) = n$ lines, and $t \cdot \Theta(n^{4/3}/t^{2/3}) = \Theta(n^{4/3}t^{1/3})$ incidences. Add $n$ vertical lines, one through each of the $n$ original points in the $xy$-plane. This turns all the points into joints, and we now have $2n$ lines. Substituting $t = m/n$, the number of incidences is

$$\Theta(n^{4/3}t^{1/3}) = \Theta(m^{1/3}n),$$

as asserted. □

**Proof of Theorem 11.** By the discussion following the theorem statement, we may assume that $m \geq n$ and focus on establishing the first bound $I(P, L) = O(m^{1/3}n)$.

We prove, using induction on the number of lines, that $I(P, L) \leq Am^{1/3}n$, for some sufficiently large absolute constant $A$, whose choice will be dictated by various constraints imposed during the proof. By choosing $A \geq n_0^{4/3}$, for some constant $n_0$ (which we will also fix later), we guarantee that the theorem is true for $n \leq n_0$, since, for $n \leq n_0$, the number of incidences $I$ satisfies

$$I \leq mn = m^{2/3} \cdot m^{1/3}n < n^{4/3} \cdot m^{1/3}n \leq n_0^{4/3} \cdot m^{1/3}n \leq Am^{1/3}n.$$

To establish the induction step, we assume that the theorem holds for each pair of sets $L', P'$ satisfying its assumptions, with $|L'| < n$. Let $L$ be a set of $n$ lines and $P$ a set of $m$ points satisfying the assumptions of the theorem, and suppose to the contrary that $I(P, L) > Am^{1/3}n$. As noted above, we may assume that $m \geq n$.

For $a \in P$, let $\mu(a)$ denote the multiplicity of $a$, which is the number of lines of $L$ incident to $a$. Similarly, for $\ell \in L$, let $\nu(\ell)$ denote the multiplicity of $\ell$, which is the number of points of $P$ lying on $\ell$.

**Pruning.** We begin by applying the following pruning process. Put $\nu = cm^{1/3}$, for some (sufficiently large) constant $c \ll A$ which we will fix later. As long as there exists a line $\ell \in L$ with $\nu(\ell) < \nu$, we remove $\ell$ from $L$, but do not remove any point incident to $\ell$. We keep repeating this step (without changing $\nu$), until each of the surviving lines has multiplicity at least $\nu$. However, if, during the pruning process, some point $a$ loses $\lfloor \mu(a)/2 \rfloor$ incident lines, we remove $a$ from $P$. This decreases the multiplicity of some lines, and we use the new multiplicities in the test for pruning further lines, but we keep using the original threshold $\nu$.

When we delete a line $\ell$, we lose at most $\nu$ incidences with surviving points. When a point $a$ is removed, the number of current incidences with $a$ is smaller than or equal to twice the number of incidences with $a$ that have already been removed. Hence, the total number of incidences that were lost during the pruning process is at most $3n\nu = 3cm^{1/3}n$. Thus, we are left with a subset $P_1$ of the points and with a subset $L_1$ of the lines, so that each $\ell \in L_1$ contains at least $\nu = cm^{1/3}$ points of $P_1$, and each point $a \in P_1$ is incident to at least three lines of $L_1$ (the latter is an immediate consequence of the rule for pruning a point).

**Sampling.** Draw a random sample $P_1^s$ of $P_1$ by choosing each point independently with probability $t$, for $t < 1$ a small constant, whose concrete value will be determined later. The expected size of $P_1^s$ is $tm$. Each line $\ell \in L_1$ contains at least $cm^{1/3}$ points of $P_1$, so the expected number of points of $P_1^s$ on $\ell$ is at least $ctm^{1/3}$. Since $m \geq n$, we can apply Chernoff's bound and the probability union bound to conclude that, with positive probability, we have $|P_1^s| \leq 2tm$ and $|\ell \cap P_1^s| \geq \frac{1}{2}ctm^{1/3}$ for every line $\ell \in L$. We assume that our sample does indeed have these properties.

**The polynomial $p$.** Now construct, using Proposition 5, a square-free trivariate polynomial $p$ which vanishes on $P_1^s$, whose degree is at most the smallest integer $d$ satisfying $\binom{d+3}{3} \geq |P_1^s| + 1$, so

$$d \leq \lceil (6|P_1^s|)^{1/3} \rceil \leq (12tm)^{1/3} + 1 < 3(tm)^{1/3},$$

for $n_0$ sufficiently large and $t$ sufficiently small (recall that $m \geq n > n_0$).

By choosing $c > 6/t^{2/3}$ we guarantee that $|\ell \cap P_1^s| \geq \frac{1}{2}ctm^{1/3} > d$ for each $\ell \in L_1$. Hence $p$ vanishes identically on each line of $L_1$. Hence, in particular, $p$ vanishes at all the points of $P_1$. Moreover, for $t$ sufficiently small, we will also have $c > 15t^{1/3}$, which guarantees that each line of $L_1$ is incident to at least $5d$ points of $P_1$.

Finally, we choose $t$ sufficiently small so as to guarantee that $4d^2 < n/2$. That is, we require that $4 \cdot 9t^{2/3}m^{2/3} < n/2$, or that $m < \dfrac{1}{72^{3/2}t}n^{3/2}$. This will indeed hold if $t$ is sufficiently small, in view of Theorem 9.

As in the proof of Theorem 9, the analysis will next argue that each line of $L_1$ must contain either at least $d$ critical points or at least $3d$ flat points of $p$, so each line is either a critical line or a flat line. However, in order to apply Proposition 8, we will first need to get rid of any linear factors of $p$, which we do using the following technique.

Clearly $p$ can have at most $d$ linear factors; i.e., $p$ can vanish identically on at most $d$ planes $\pi_1, \ldots, \pi_k$, for $k \leq d$. We factor out all the linear factors from $p$, and let $\tilde{p}$ denote the resulting polynomial, which is a square-free polynomial without any linear factors, of degree at most $d$. Let $L_2 \subseteq L_1$ (resp., $P_2 \subseteq P_1$) denote the set of those lines of $L_1$ (resp., points of $P_1$) which are not contained in any of the vanishing planes $\pi_i$. Put $L_2' = L_1 \setminus L_2$ and $P_2' = P_1 \setminus P_2$.

For each line $\ell \in L_2$, $\tilde{p}$ vanishes identically on $\ell$, and at most $d$ points of $P_1 \cap \ell$ lie in the planes $\pi_i$. Hence, $\ell$ contains at least $4d$ points of $P_2$, and, arguing as in the preceding proof, each of these points is either critical or flat for $\tilde{p}$. Hence, either at least $d$ of these points are critical, and then $\ell$ is a critical line for $\tilde{p}$, or at least $3d$ of these points are flat, and then $\ell$ is a flat line for $\tilde{p}$. Applying Propositions 3 and 8, the overall number of lines in $L_2$ is therefore at most
$$d(d-1) + d(3d-4) < 4d^2 < \frac{n}{2}.$$

We now apply the induction hypothesis to the sets $L_2$ and $P_2$, with $4d^2$ as the bound on the size of $|L_2|$. For this we need to argue that no plane contains more than $4bd^2$ points of $P_2$, which we do exactly as in the proof of Theorem 9. Hence, by induction, we have
$$I(P_2, L_2) = I(P_2, L_1) \leq Am^{1/3}\frac{n}{2} = \frac{1}{2}Am^{1/3}n.$$

Next, we bound the number of incidences between $P_2'$ and $L_1$, namely, between the points contained in the vanishing planes and all the lines of $L_1$. To do so, we iterate over the planes, say, in the order $\pi_1, \ldots, \pi_k$. For each plane $\pi_i$ in turn, we process the points and lines contained in $\pi_i$ and then remove them from further processing on subsequent planes. Let $m_{\pi_i}$ denote the number of surviving points of $P_2'$ which lie on $\pi_i$, and let $n_{\pi_i}$ denote the number of surviving lines of $L_2'$ contained in $\pi_i$. The number of incidences between these points and lines is [13]
$$O\left(m_{\pi_i}^{2/3}n_{\pi_i}^{2/3} + m_{\pi_i} + n_{\pi_i}\right).$$

We now remove all these $m_{\pi_i}$ points and $n_{\pi_i}$ lines, and repeat this analysis for the other planes $\pi_j$ on which $p$ vanishes. Summing over all the $k \leq d$ vanishing planes, we count
$$O\left(\sum_{i=1}^{k}\left(m_{\pi_i}^{2/3}n_{\pi_i}^{2/3} + m_{\pi_i} + n_{\pi_i}\right)\right),$$

incidences.

Note, though, that not all incidences are counted. An incidence between a point $a \in P_2'$ and a line $\ell \in L_1$ can only be detected within the first plane $\pi_j$ containing $a$. For this, $\ell$ must be contained in $\pi_j$ and in no previously processed plane. Clearly, $\ell$ cannot lie in any previous plane, because then $a$ would also have to lie in that plane, contrary to assumption. However, it is possible that $\ell$ is not contained in $\pi_j$ but in some later plane, or that $\ell \in L_2$ and is therefore not contained in any plane. In these cases $a$ is the unique intersection point of $\ell$ with $\pi_j$, so the number of incidences that we miss on each line of $L_1$ is at most $d$ (the number of times it intersects the vanishing planes). Since we assumed that each line of $L_1$ contains at least $5d$ points, the number of missed incidences of this kind is at most one fifth of the incidences that have been counted.

By assumption, no plane can contain more than $bn$ points of $P$. Hence, the overall number of incidences between the points that lie in the vanishing planes and *all* the lines of $L_1$ is at most (where $m_0 = |P_2'|$)

$$I(P_2', L_1) = O\left(nd + \sum_{i=1}^{k}\left(m_{\pi_i}^{2/3}n_{\pi_i}^{2/3} + m_{\pi_i} + n_{\pi_i}\right)\right) = O\left(m_0 + nd + n^{1/3} \cdot \left(\sum_{i=1}^{k} m_{\pi_i}^{1/3} n_{\pi_i}^{2/3}\right)\right) =$$

$$O\left(m_0 + nd + n^{1/3}\left(\sum_{i=1}^{k} m_{\pi_i}\right)^{1/3}\left(\sum_{i=1}^{k} n_{\pi_i}\right)^{2/3}\right) = O\left(m_0 + nd + m_0^{1/3} n^{2/3} n^{1/3}\right) =$$

$$O\left(m + m^{1/3}n\right),$$

and we write this bound as $B(m + m^{1/3}n)$ for an appropriate absolute constant $B$. Adding the bound $I(P_2, L_1) \le \frac{1}{2}Am^{1/3}n$, we get that

$$I(P_1, L_1) \le \frac{1}{2}Am^{1/3}n + B(m + m^{1/3}n).$$

Since $m = O(m^{1/3}n)$ for $m = O(n^{3/2})$, it follows that

$$I(P_1, L_1) \le \frac{1}{2}Am^{1/3}n + B'm^{1/3}n,$$

for another absolute constant $B'$. Adding the incidences discarded in the initial pruning step we finally get that

$$I(P, L) \le \frac{1}{2}Am^{1/3}n + B'm^{1/3}n + 3cm^{1/3}n$$

Choosing $A > 2(B' + 3c)$, we get $I(P, L) \le Am^{1/3}n$, which contradicts our assumptions on $L$ and $P$, and thus establishes the induction step, and completes the proof of the theorem. □

**Remark.** The hardest case of Theorem 11 is when $m = \Theta(n^{3/2})$. If $m \ll n^{3/2}$ (but is still at least $n$) we can construct a polynomial $p$ which vanishes at *all* the points of $P_1$, avoiding the need to construct the sample $P_1^s$; in this case the degree of $p$ satisfies $d < 3m^{1/3}$. As is easily verified, with the exception of one constraint, all the constraints on $d$, in relation to the other parameters, can be satisfied with this larger value of $d$, with an appropriate choice of the other parameters. The problematic constraint is $4d^2 < n/2$, whose enforcement does require sampling. However, when $m \ll n^{3/2}$, we have $4d^2 < 36m^{2/3} \ll n$, so this constraint is also satisfied. This allows us to apply the induction hypothesis to $P_2$ and $L_2$, and thus carry out the proof without sampling, resulting in an even simpler proof.

It is interesting to note that the proof technique also yields the following result.

**Proposition 12** *Let $p$ be a square-free trivariate polynomial of degree $d$ with no linear factors, and let $Z$ denote the zero set of $p$. Let $L$ be a set of $n$ arbitrary lines, and let $P$ be the set of all points in $Z$ which lie on at least three lines of $L$. Then $|P|$ and $I(P, L)$ are both $O(nd + d^3)$.*

**Proof.** Any line $\ell \in L$ which is not fully contained in $Z$ (a "crossing" line) can have at most $d$ incidences with the points of $P$, so it suffices to consider lines fully contained in $Z$. Each such line $\ell$ is either a critical line for $p$, or a flat line for $p$, or an "ordinary" line, namely, neither critical nor flat. Let $L_1$ denote the subset of critical and flat lines in $L$.

By Propositions 3 and 8, we have $|L_1| < 4d^2$. We can apply Theorems 9 and 11 to $L_1$ and to the subset $P_1$ of $P$ consisting of those points incident to at least three lines of $L_1$. Condition (i) of the theorems holds because no plane can contain more than $d$ lines of $L_1$ (or else $p$ would have vanished on such a plane, and thus have a linear factor, contrary to assumption). This implies that the number of points of $P_1$ on any fixed plane is at most $O(d^2)$, arguing exactly as in the preceding proofs. Hence, $|P_1| = O(d^3)$ and, consequently, $I(P_1, L_1) = O(d^3)$ too.

A point $a \in P \setminus P_1$ has at most two incidences with the lines of $L_1$, and we charge these incidences to the incidence(s) of $a$ with the other (ordinary or crossing) lines of $L$ (by assumption, there has to exist at least one such incidence).

Since we have already handled the crossing lines, it remains to bound the number of incidences with ordinary lines. Such a line $\ell$ contains fewer than $d$ critical points and fewer than $3d$ flat points, for a total of at most $4d$ incidences with such points. Any other point $a \in P$ incident to $\ell$ is incident to at most one additional line contained in $Z$ (necessarily an ordinary line). Thus $a$ can have at most two incidences with ordinary lines, and thus at least one incidence with a crossing line. Since the overall number of incidences of the latter kind is at most $nd$, the number of incidences of the former kind is at most $2nd$.

Since we have accounted for all possible incidences, the asserted bounds on $|P|$ and on $I(P, L)$ follow. $\square$

**Corollary 13** *Let $L$ be a set of $n$ lines and $P$ a set of points in 3-space which satisfy the conditions of Theorem 11. Then, for any $k \geq 1$, the number $M_{\geq k}$ of points of $P$ incident to at least $k$ lines of $L$ satisfies*

$$M_{\geq k} = \begin{cases} O\left(\dfrac{n^{3/2}}{k^{3/2}}\right) & \text{for } k \leq n^{1/3}, \\ O\left(\dfrac{n^2}{k^3} + \dfrac{n}{k}\right) & \text{for } k > n^{1/3}, \end{cases}$$

and the number $I_{\geq k}$ of incidences between these points and the lines of L satisfies

$$I_{\geq k} = \begin{cases} O\left(\dfrac{n^{3/2}}{k^{1/2}}\right) & \text{for } k \leq n^{1/3}, \\ O\left(\dfrac{n^2}{k^2} + n\right) & \text{for } k > n^{1/3}. \end{cases}$$

**Proof.** Write $m = M_{\geq k}$ for short. We clearly have $I(P, L) \geq km$. Theorem 11 then implies $km = O(m^{1/3}n)$, or $m = O((n/k)^{3/2})$. If $k > n^{1/3}$ we use the other bound $km = O(m^{2/3}n^{2/3} + m + n)$ to deduce that $m = O(n^2/k^3 + n/k)$ (which is in fact an equivalent statement of the classical Szemerédi-Trotter bound). The corresponding bounds for $I_{\geq k}$ follow immediately from Theorem 11. □

## 4.1 A proof of Bourgain's lemma

The work on Kakeya's problem has inspired Bourgain to make the following conjecture, which has been settled in the affirmative by Guth and Katz [7]. The proof technique of Theorem 11 can be adapted to yield an alternative, and, in our opinion, simpler proof of Bourgain's conjecture.

**Proposition 14** *Let L be a set of n lines and P be a set of points in $\mathbb{R}^3$, such that (i) each line is incident to at least $n^{1/2}$ points of P, and (ii) no plane contains more than $n^{1/2}$ lines of L. Then $|P| = \Omega(n^{3/2})$.*

**Proof.** Set $\nu = \frac{1}{2}n^{1/2}$. Call a point $a \in P$ *light* (resp., *heavy*) if $a$ is incident to at most two (resp., at least three) lines of L. Call a line $\ell \in L$ *light* (resp., *heavy*) if it contains fewer than (resp., at least) $\nu$ heavy points.

If at least half of the lines are light, we get at least $\frac{1}{4}n^{3/2}$ incidences with light points. Since each light point can be incident to at most two lines, we must have

$$|P| \geq \frac{1}{8}n^{3/2},$$

so the asserted bound holds in this case.

Assume then that at least half of the lines of L are heavy, and ignore all the light lines. Let $L_1$ denote the set of the remaining, heavy lines, and set $n_1 = |L_1| \geq n/2$. Note that, in removing the light lines, some heavy points may have become light with respect to $L_1$, and some points may have "vanished" altogether, in the sense that they are not incident to any line of $L_1$; we ignore this latter kind of points, denote by $P_1$ the subset of surviving points (both light and heavy), and set $m = |P_1|$.

Construct a square-free polynomial $p$ which vanishes at all the points of $P_1$, using Proposition 5; the degree $d$ of $p$ satisfies $d \leq cm^{1/3}$, where $c \approx 6^{1/3}$. We may assume that $\nu > 10d$; if not, we have $\frac{1}{2}n^{1/2} \leq 10cm^{1/3}$, or $|P| \geq m \geq c'n^{3/2}$, for $c' = (1/(20c))^3$, implying the asserted bound. Similarly, we will also assume that $32d^2 < n$; again, if this were not the case, we would get the asserted bound $m = \Omega(n^{3/2})$.

Arguing as in the preceding proofs, $p$ vanishes identically on every line of $L_1$.

Before continuing, in preparation for the application of Proposition 8, we get rid of all the linear factors of $p$, if any. For this, we note that $p$ can have at most $d$ linear factors, so it can vanish identically on at most $d$ planes. By assumption, each such plane can contain at most $n^{1/2}$ lines of $L_1$, so the overall number of lines contained in the union of these planes is at most $dn^{1/2}$, which is smaller than $n/4$, by the preceding assumption. We therefore conclude that at least $n_1 - n/4 \geq n/4$ lines of $L_1$ are not contained in any vanishing plane. Denote by $L_2$ the subset of these surviving lines, and by $P_2$ the subset of points of $P_1$ which do not lie in any vanishing plane.

Next, we factor out all the linear factors from $p$, and let $\tilde{p}$ denote the resulting polynomial, which is a square-free polynomial without any linear factors, of degree at most $d$. Clearly, $\tilde{p}$ vanishes identically on every line of $L_2$. Moreover, any such line $\ell$ has at most $d$ points that lie on the union of the vanishing planes, so it still contains at least $\nu - d \geq \frac{1}{2}\nu + 4d$ points of $P_2$.

Any point $a \in P_2$ is either a *light* point, if it is incident to at most two lines of $L_2$, or a *critical* point of $\tilde{p}$, if it is incident to three non-coplanar lines of $L_2$, or a *flat* point of $\tilde{p}$, if it is incident to at least three lines of $L_2$, all coplanar.

Fix a line $\ell \in L_2$. Since $\ell$ contains at least $\frac{1}{2}\nu + 4d$ points of $P_2$, it must contain either (i) at least $d$ critical points, in which case $\ell$ is a critical line for $\tilde{p}$, or (ii) at least $3d$ flat points, in which case $\ell$ is a flat line for $\tilde{p}$, or (iii) at least $\frac{1}{2}\nu$ light points, in which case we call it a *light* line (with respect to $P_2$ and $L_2$).

By Propositions 3 and 8, the number of critical (resp., flat) lines for $\tilde{p}$ is at most $d(d-1)$ (resp., $d(3d-4)$). Hence the number of these two kinds of lines is at most $d^2 + 3d^2 = 4d^2 < n/8$. Since $|L_2| \geq n/4$, at least $n/8$ lines of $L_2$ are light, so each of them is incident to at least $\frac{1}{2}\nu$ light points of $P_2$. We therefore get at least $\frac{1}{32}n^{3/2}$ such incidences, and since each light point participates in at most two incidences, the number of light points is at least $\frac{1}{64}n^{3/2}$, and the asserted bound follows. This completes the proof. □

## 5 Further implications

As mentioned in the introduction, incidences between points and lines in 3-space have been previously studied by Sharir and Welzl [11], who have obtained several weaker bounds, which can now be improved, using the new results of this paper. For example, the following result improves Theorem 4.1 of [11].

**Theorem 15** *Let $P$ be a set of $m$ points and $L$ a set of $n$ lines in $\mathbb{R}^3$. For each $p \in P$ define the* plane cover $c_p(L)$ *of $p$ to be the smallest number of planes which cover all the lines of $L$ incident to $p$, and put $I_c(P, L) = \sum_{p \in P} c_p(L)$. Then we have*

$$I_c(P, L) = O(m^{1/2}n^{3/4} + m + n).$$

**Proof.** We apply the same partitioning scheme used in [11], which decomposes the input into $O(r^2)$ subproblems, each involving at most $m/r$ points of $P$ and at most $n/r^2$ lines of $L$. We note that, within each subproblem, a point is either a joint, or its plane cover is 1. Applying Theorem 10 to each subproblem, the sum of the plane covers of the joints, which is upper bounded by the number of their incidences, is $O((m/r)^{1/3}n/r^2)$. The sum of the plane covers of the other points is $O(m/r)$. Summing over all subproblems, as in [11], we obtain

$$I_c(P, L) = O(r^2) \cdot O\left(\left(\frac{m}{r}\right)^{1/3} \frac{n}{r^2} + \frac{m}{r}\right) = O\left(mr + \frac{m^{1/3}n}{r^{1/3}}\right).$$

We choose $r = n^{3/4}/m^{1/2}$, and note that $1 \le r \le m$ when $n^{1/2} \le m \le n^{3/2}$. For $m$ outside this range, the bound is easily seen to be $O(m + n)$, and within this range the bound is $O(m^{1/2}n^{3/4})$. This completes the proof. $\square$

**Remarks.**
(1) The bound $O(m^{1/2}n^{3/4})$ is strictly smaller than either of the bounds $O(m^{1/3}n)$ or $O(m^{2/3}n^{2/3})$ for every $m$ strictly between $n^{1/2}$ and $n^{3/2}$. This means that, even for joints, measuring incidences by plane covers may yield a smaller value than the actual number of incidences. This is for example the case in the lower bound constructions in Theorems 10 and 11, where $I_c(P, L)$ is only $O(m + n)$.

(2) We do not know whether the bound on $I_c(P, L)$ is tight in the worst case.

(3) Another problem studied in [11] involves incidences between points and equally-inclined lines in $\mathbb{R}^3$. In view of the improvement just obtained, the upper bound on $I(P, L)$ improves in this case to

$$O\left(\min\left\{m^{3/4}n^{1/2}\log^{1/2} m, \ m^{1/2}n^{3/4}\right\} + m + n\right);$$

see [11] for details concerning the first term in the above expression. A construction in [11] gives a lower bound of $\Omega(m^{2/3}n^{1/3})$, for $m = \Omega(n^{3/4})$. It remains to close the gap between the upper and lower bounds.

**Acknowledgments.** We wish to thank Roel Apfelbaum, Boris Aronov, János Pach, Shakhar Smorodinsky, and Emo Welzl for helpful discussions concerning the problems studied in this paper.

## A  The vanishing of the second fundamental form

For the sake of completeness, we provide here a short proof of the property used in the proof of Proposition 7. That is:

**Claim:** If the second fundamental form of the zero set $Z$ of a trivariate polynomial $p$ vanishes at all the regular points of $p$ then $Z$, locally near any regular point, is a portion of a plane.

**Proof.** Recall that the second fundamental form of the zero set $Z$ of $p$ at a regular point $a$ is defined as $A du^2 + 2B du dv + C dv^2$, where $\mathbf{x} = \mathbf{x}(u,v)$ is a parametrization of $Z$ (locally near $a$), and
$$A = \mathbf{x}_{uu} \cdot \mathbf{n}, \qquad B = \mathbf{x}_{uv} \cdot \mathbf{n}, \qquad C = \mathbf{x}_{vv} \cdot \mathbf{n},$$
where $\mathbf{n} = \nabla p(a)/\|\nabla p(a)\|$ is the unit normal to $Z$ at $a$.

For any $u, v$, $\mathbf{x}_u$ and $\mathbf{x}_v$ are vectors in the tangent plane to $Z$ at $\mathbf{x}(u,v)$, and thus satisfy
$$\mathbf{x}_u \cdot \mathbf{n} = \mathbf{x}_v \cdot \mathbf{n} \equiv 0.$$

We now differentiate these equations with respect to $u$ and $v$. For example, differentiating the first equation with respect to $u$ yields
$$\mathbf{x}_{uu} \cdot \mathbf{n} + \mathbf{x}_u \cdot \mathbf{n}_u \equiv 0.$$

The first term vanishes because the second fundamental form vanishes, so we have $\mathbf{x}_u \cdot \mathbf{n}_u \equiv 0$. Similar relations result from the other differentiations, and we get

$$\mathbf{x}_u \cdot \mathbf{n}_u = \mathbf{x}_v \cdot \mathbf{n}_u = \mathbf{x}_u \cdot \mathbf{n}_v = \mathbf{x}_v \cdot \mathbf{n}_v \equiv 0.$$

In other words, $\mathbf{n}_u$ and $\mathbf{n}_v$ are both orthogonal to the tangent plane of $Z$ and thus must both be parallel to $\mathbf{n}$. However, since $\mathbf{n}$ is of unit length, we have $\mathbf{n} \cdot \mathbf{n} \equiv 1$, and differentiating this equation yields

$$\mathbf{n}_u \cdot \mathbf{n} = \mathbf{n}_v \cdot \mathbf{n} \equiv 0.$$

Thus, if $\mathbf{n}_u$ is parallel to $\mathbf{n}$ it must be zero, and similarly for $\mathbf{n}_v$. That is, the vector function $\mathbf{n}(u, v)$ is constant in a neighborhood of $a$, so $Z$ must be a part of a plane near every regular point. □